\RequirePackage{ifpdf}

\newcommand{\commondocopts}{letterpaper,aps,prl,10pt,superscriptaddress,showpacs,floats,nofootinbib,twocolumn,lengthcheck}
\ifpdf
  \documentclass[\commondocopts,pdftex]{revtex4-1}
  \usepackage{cmap}
\else
  \documentclass[\commondocopts,ps2pdf]{revtex4-1}
\fi
\usepackage[latin1]{inputenc}
\usepackage[T1]{fontenc}
\usepackage{subfigure}
\usepackage{amsmath,amssymb,mathbbol}

\usepackage{float}

\usepackage{graphicx,color}
\usepackage{hyperref}

\newcommand{\titlestr}{Probing time-ordering in two-photon double ionization of helium on the attosecond time scale}

\definecolor{rltred}{rgb}{0.75,0,0}
\definecolor{rltgreen}{rgb}{0,0.5,0}
\definecolor{rltblue}{rgb}{0,0,0.75}
\definecolor{rltdblue}{rgb}{0.2,0.2,0.65}
\definecolor{rltdred}{rgb}{0.65,0.2,0.2}
\definecolor{forestgreen}{rgb}{0.13,0.54,0.13}

\hypersetup{colorlinks,%
hypertexnames=true,%
pdfauthor={R. Pazourek, S. Nagele, J. Burgd\"orfer},%
pdftitle={\titlestr},%
pdfview=FitH,%
pdfstartview=FitH,%
pdfpagemode=UseNone,%
bookmarksopen=true,%
bookmarksnumbered=true,%
pdfhighlight=/I,%
linkcolor=rltred,%
citecolor=rltgreen,%
linktocpage=true}

\begin{document}
\renewcommand{\equationautorefname}{Eq.}
\renewcommand{\figureautorefname}{Fig.}

\newcommand{\etal}{\emph{et~al.}\ }
\newcommand{\subscr}[1]{_{\scriptstyle{\mathrm{#1}}}}

\newcommand{\Int}{\int\limits}
\newcommand{\IInt}{\iint\limits}
\newcommand{\IIInt}{\iiint\limits}
\newcommand{\IIIInt}{\iiiint\limits}
\newcommand{\sumint}[1]{\sum\mkern-25mu\Int_{#1}\;}
\newcommand{\ui}{\mathrm{i}}
\newcommand{\ue}{\mathrm{e}}
\newcommand{\Vol}{\operatorname{vol}}
\newcommand{\R}{\mathbb{R}}
\newcommand{\Z}{\mathbb{Z}}
\newcommand{\N}{\mathbb{N}}
\newcommand{\cA}{\mathcal{A}}
\newcommand{\E}{\mathcal{E}}
\newcommand{\setsep}{ \;\; | \;\;}

\newcommand{\COMMENT}[1]{\textcolor{rltred}{\textbf{\textsc{#1}}}}
\newcommand{\ie}{i.e., }
\newcommand{\Schro}{Schr\"o\-din\-ger }
\newcommand{\eg}{e.g.\@ }
\newcommand{\cf}{cf.~}

\newcommand{\expval}[1]{\langle#1\rangle}
\newcommand{\abs}[1]{\left|#1\right|}
\newcommand{\norm}[1]{\left\|#1\right\|}

\newcommand{\bra}[1]{\langle#1|}
\newcommand{\ket}[1]{|#1\rangle}
\newcommand{\braket}[2]{\langle#1|#2\rangle}
\newcommand{\braOket}[3]{\langle#1|#2|#3\rangle}
\newcommand{\expt}[1]{\langle#1\rangle}

\newcommand{\doubD}{{\mathord{\buildrel{\lower3pt\hbox{$\scriptscriptstyle\leftrightarrow$}}\over {\bf D}}}}
\newcommand{\unitv}[1]{\mathbf{\hat{#1}}}

\newcommand{\refeq}[1]{\hyperref[#1]{\equationautorefname~(\ref*{#1})}}

\newcommand{\cvec}[1]{\mathbf{#1}}

\newcommand{\op}[1]{\mathrm{\hat{#1}}}
\newcommand{\vecop}[1]{\cvec{\hat{#1}}}
\newcommand{\eqcomma}{\,,}
\newcommand{\eqstop}{\,.}
\newcommand{\ed}{\,}

\newcommand{\ddE}{\frac{\partial}{\partial E}}
\newcommand{\ddt}{\frac{\partial}{\partial t}}
\newcommand{\dx}{\dd x}
\newcommand{\dt}{\dd t}
\newcommand{\dr}{\dd r}
\newcommand{\dw}{\dd\omega}
\newcommand{\dwb}{\dd\bar{\omega}}
\newcommand{\dW}{\dd\Omega}
\newcommand{\dE}{\dd E}
\newcommand{\dk}{\dd k}
\newcommand{\dd}{\mathrm{d}}
\newcommand{\Dt}{\Delta t}
\newcommand{\hw}{\hbar\omega}

\newcommand{\domega}{\,\mathrm{d}\Omega}
\newcommand{\dtheta}{\,\mathrm{d}\theta}
\newcommand{\dphi}{\,\mathrm{d}\varphi}

\newcommand{\sube}{{\mathrm{e}}}
\newcommand{\subn}{{\mathrm{n}}}

\newcommand{\csph}{{\mathcal{Y}}}
\newcommand{\sph}[2]{{\mathrm{Y}_{#2}^{(#1)}}}
\newcommand{\sphcmplx}[2]{{\mathrm{Y}_{#2}^{*(#1)}}}
\newcommand{\rensph}[2]{{\mathrm{C}_{#2}^{(#1)}}}
\newcommand{\renredsph}[1]{{\mathrm{C}^{(#1)}}}
\newcommand{\submax}{\mathrm{max}}
\newcommand{\Lmax}{L_\submax}
\newcommand{\lonemax}{l_{1,\submax}}
\newcommand{\ltwomax}{l_{2,\submax}}

\newcommand{\cmfs}{\,\mathrm{cm}^4\mathrm{s}}
\newcommand{\ev}{\,\mathrm{eV}}
\newcommand{\eV}{\ev}
\newcommand{\au}{\,\mathrm{a.u.}}
\newcommand{\nm}{\,\mathrm{nm}}
\newcommand{\Wcm}{\,\mathrm{W}/\mathrm{cm}^2}
\newcommand{\as}{\,\mathrm{as}}
\newcommand{\fs}{\,\mathrm{fs}}
\newcommand{\He}{\mathrm{He}}
\newcommand{\Hep}{\He^+}
\newcommand{\Hepp}{\He^{++}}
\newcommand{\kone}{{\cvec{k}_1}}
\newcommand{\ktwo}{{\cvec{k}_2}}

\newcommand{\ti}{{t_\mathrm{(i)}}}
\newcommand{\tii}{{t_\mathrm{(ii)}}}
\newcommand{\tiii}{{t_\mathrm{(iii)}}}
\newcommand{\tcor}{{t_\mathrm{cor}}}
\newcommand{\Tp}{{T_{\mathrm{p}}}}
\newcommand{\Teff}{{T_{\mathrm{eff}}}}
\newcommand{\Tramp}{{T_{\mathrm{ramp}}}}
\newcommand{\Tfull}{{T_{\mathrm{full}}}}
\newcommand{\Ene}{{E_\mathrm{ne}}}
\newcommand{\Eeqs}{{E_\mathrm{eq}}}
\newcommand{\Eexc}{{E_\mathrm{exc}}}
\newcommand{\Etot}{{E_\mathrm{tot}}}
\newcommand{\Efseq}{{E^\mathrm{seq}_1}}
\newcommand{\Esseq}{{E^\mathrm{seq}_2}}
\newcommand{\Efsseq}{{E^\mathrm{seq}_{1,2}}}
\newcommand{\PDI}{{P^{DI}}}
\newcommand{\PDIs}{{P^{DI}_\mathrm{seq}}}
\newcommand{\PDIns}{{P^{DI}_\mathrm{nonseq}}}
\newcommand{\Pasym}{\mathcal{A}}
\newcommand{\DE}{\Delta E}
\newcommand{\G}{\mathcal{G}}
\newcommand{\tews}{t_{\scriptstyle\mathrm{EWS}}}
\newcommand{\tEWS}{t_{\scriptstyle\mathrm{EWS}}}
\newcommand{\tewsdi}{t_{\scriptstyle\mathrm{EWS}}^\mathrm{\scriptstyle{{DI}}}}
\newcommand{\Tewsdi}{T_{\scriptstyle\mathrm{EWS}}^\mathrm{\scriptstyle{{DI}}}}
\newcommand{\tewsdii}{t_{\scriptstyle\mathrm{EWS,}j}^\mathrm{\scriptstyle{{DI}}}}
\newcommand{\tewsdione}{t_{\scriptstyle\mathrm{EWS,1}}^\mathrm{\scriptstyle{{DI}}}}
\newcommand{\tewsditwo}{t_{\scriptstyle\mathrm{EWS,2}}^\mathrm{\scriptstyle{{DI}}}}
\newcommand{\tewssi}{t_{\scriptstyle\mathrm{EWS}}^\mathrm{\scriptstyle{{(\gamma)}}}}
\newcommand{\tewssijk}{t_{\scriptstyle\mathrm{EWS},j}^{\scriptstyle{{(\gamma_m)}}}}
\newcommand{\tewssii}{t_{\scriptstyle\mathrm{EWS,}j}^\mathrm{\scriptstyle{{(\gamma)}}}}
\newcommand{\tewssione}{t_{\scriptstyle\mathrm{EWS,1}}^\mathrm{\scriptstyle{{(\gamma)}}}}
\newcommand{\tewssitwo}{t_{\scriptstyle\mathrm{EWS,2}}^\mathrm{\scriptstyle{{(\gamma)}}}}
\newcommand{\tewsav}{\expval{t_{\scriptstyle\mathrm{EWS}}}}
\newcommand{\tewscoul}{t_{\scriptstyle\mathrm{EWS}}^\mathrm{C}}
\newcommand{\tclc}{t_{\scriptstyle\mathrm{CLC}}}
\newcommand{\tCLC}{t_{\scriptstyle\mathrm{CLC}}}
\newcommand{\tclci}{t_{\scriptstyle\mathrm{CLC,}j}}
\newcommand{\tislc}{t_{\scriptstyle\mathrm{ISLC}}}
\newcommand{\tfslc}{t_{\scriptstyle\mathrm{FSLC}}}
\newcommand{\tst}{t_{\scriptstyle\mathrm{S}}}
\newcommand{\tSG}{t_{\scriptstyle\mathrm{S}}^\G}
\newcommand{\tSSFA}{\delta t^{\scriptstyle(2\gamma,2e)}}
\newcommand{\tSSFAi}{\delta t_{\scriptstyle{j}}^{\scriptstyle(2\gamma,2e)}}
\newcommand{\tSGi}{t_{\scriptstyle\mathrm{S,}j}^\G}
\newcommand{\tGo}{t^{(1\gamma)}_\G}
\newcommand{\tGt}{t^{(2\gamma)}_\G}
\newcommand{\tGtj}{t^{(2\gamma)}_{\G,j}}
\newcommand{\tSdi}{t_{\scriptstyle\mathrm{S}}^\mathrm{\scriptstyle{{DI}}}}
\newcommand{\tSdii}{t_{\scriptstyle\mathrm{S,}j}^\mathrm{\scriptstyle{{DI}}}}
\newcommand{\tSdione}{t_{\scriptstyle\mathrm{S,}1}^\mathrm{\scriptstyle{{DI}}}}
\newcommand{\tSditwo}{t_{\scriptstyle\mathrm{S,}2}^\mathrm{\scriptstyle{{DI}}}}
\newcommand{\tstainf}{t_{\scriptstyle\mathrm{S,a\rightarrow\infty}}}
\newcommand{\tstcor}{t_{\scriptstyle\mathrm{S,c}}}
\newcommand{\tprop}{t_\mathrm{prop}}
\newcommand{\tbh}{t_{\scriptstyle\mathrm{BH}}}
\newcommand{\twp}{t_{\scriptstyle\mathrm{WP}}}
\newcommand{\twploc}{t_{\scriptstyle\mathrm{WP}}^{\scriptstyle\mathrm{loc}}}
\newcommand{\twpf}{\mathcal{T}}
\newcommand{\twpfloc}{\mathcal{T}^{\scriptstyle\mathrm{loc}}}

\newcommand{\dtuc}{\expval{\Delta t}\subscr{uc}}

\newcommand{\tCoul}{t_{\scriptstyle\textrm{Coul}}}
\newcommand{\tCoulcl}{t_{\scriptstyle\textrm{Coul}}^{\scriptstyle\textrm{cl.}}}
\newcommand{\tcc}{t_{\scriptstyle\textrm{cc}}}
\newcommand{\tCLCcl}{t_{\scriptstyle\textrm{CLC}}^{\scriptstyle\textrm{cl.}}}
\newcommand{\tclcmisha}{t_{\scriptstyle\textrm{CLC}}^{\scriptstyle\textrm{Ivanov}}}
\newcommand{\tclctq}{t_{\scriptstyle\textrm{CLC}}^{\scriptstyle\mathcal{T}}}
\newcommand{\tclctqbet}{t_{\scriptstyle\textrm{CLC}}^{\scriptstyle{\mathcal{T} (2)}}}
\newcommand{\toptcyc}{T_{\scriptstyle\textrm{IR}}}
\newcommand{\tabsorb}{t_{{\gamma}}}
\newcommand{\cDI}{c^{\scriptstyle\textrm{DI}}}

\newcommand{\tScl}{t_{\scriptstyle\textrm{S}}^{\scriptstyle\textrm{cl.}}}
\newcommand{\tEWScl}{t_{\scriptstyle\textrm{EWS}}^{\scriptstyle\textrm{cl.}}}
\newcommand{\tEWSCcl}{t_{\scriptstyle\textrm{EWS}}^{\scriptstyle\textrm{C, cl.}}}
\newcommand{\tEWSC}{t_{\scriptstyle\textrm{EWS}}^{\scriptstyle\textrm{C}}}
\newcommand{\tEWSCdif}{t_{\scriptstyle\textrm{EWS},\Delta E}^{\scriptstyle\textrm{C}}}
\newcommand{\tewsDI}{t_{\scriptstyle\textrm{EWS}}^{\scriptstyle\textrm{DI}}}

\newcommand{\tdLC}{t_{\scriptstyle\textrm{dLC}}}
\newcommand{\tpxuv}{\tau\subscr{XUV}}
\newcommand{\tpeak}{t\subscr{peak}}
\newcommand{\tiroc}{T\subscr{IR}}

\newcommand{\psiI}{\psi_{\mathrm{I}}}
\newcommand{\iir}{I_{\scriptstyle\mathrm{IR}}}
\newcommand{\air}{A_{\scriptstyle\mathrm{IR}}}
\newcommand{\level}[3]{{}^{#1}{#2}^{\textrm{#3}}}
\newcommand{\tS}{t_{\scriptstyle\textrm{S}}}
\newcommand{\tR}{t_{\scriptstyle\textrm{R}}}
\newcommand{\Ggauss}{\G_{G}}
\newcommand{\Ggaussint}{\tilde\G_{G}}

\newcommand{\Efn}{E_{fn}}
\newcommand{\Efnw}{E_{fn\omega}}
\newcommand{\Eni}{E_{ni}}
\newcommand{\Eniw}{E_{ni\omega}}
\newcommand{\DEn}{\Delta}
\newcommand{\DEt}{E_t}
\newcommand{\Faddeeva}{w}

\title{\titlestr}

\newcommand{\itptuw}{Institute for Theoretical Physics, Vienna University of Technology, 1040 Vienna, Austria, EU}
\newcommand{\atomki}{Institute of Nuclear Research of the Hungarian Academy of Sciences (ATOMKI), 4001 Debrecen, Hungary, EU}

\author{Renate~Pazourek}
\email{renate.pazourek@tuwien.ac.at}
\affiliation{\itptuw}
\affiliation{Department of Physics and Astronomy, Louisiana State University, Baton Rouge, Louisiana 70803, USA}

\author{Stefan~Nagele}
\affiliation{\itptuw}

\author{Joachim~Burgd\"orfer}
\affiliation{\itptuw}
\affiliation{\atomki}

\date{\today}

\begin{abstract}
We show that time ordering underlying time-dependent quantum dynamics is a physical observable accessible by attosecond streaking. 
We demonstrate the extraction of time ordering for the prototypical case of time-resolved two-photon double ionization (TPDI) of helium by an attosecond XUV pulse. 
The Eisenbud-Wigner-Smith time delay for the emission of a two-electron wavepacket and the time interval between subsequent emission events can be unambiguously determined by attosecond streaking.  
The delay between the two emission events sensitively depends on the energy, pulse duration, and angular distribution of the emitted electron pair. 
Our fully-dimensional ab-initio quantum mechanical simulations provide benchmark data for experimentally accessible observables. 

\end{abstract}

\maketitle

With recent advances in the generation of new light sources, accessing real time information
of the electronic dynamics on the attosecond scale has become possible. 
One first prototypical test case was the time resolved photoelectric effect for atoms and solid surfaces \cite{CavMueUph2007,SchFieKar2010,KluDahGis2011}. 
Relative time differences between ionization from two different subshells 
initiated by a single photon of an ultrashort XUV laser pulse have been measured by attosecond pump-probe setups employing a weak infrared (IR) field as probe and a single attosecond XUV pulse (``streaking'' \cite{ItaQueYud2002,KieGouUib2004,YakBamScr2005}) or a train of attosecond pulses (``RABBIT'' \cite{PauTomBre2001,VenTaiMaq1996,Mul2002}), that trigger the photoionization, as the pump.
A fundamental question is that of ``time zero'', \ie when does the photoemission process start \cite{SchFieKar2010}. 
The Eisenbud-Wigner-Smith (EWS) time delay $\tEWS$ \cite{Eis1948,Wig1955, Smi1960,deCNus2002} that characterizes the delay in the formation of an outgoing wavepacket has evolved as one key physical observable that has become accessible \cite{SchFieKar2010,KluDahGis2011} provided that corrections due to the probing IR field are properly taken into account \cite{NagPazFei2011,PazNagBur2013,NagPazFei2012,NagPazWai2014,DahHuiMaq2012,DahCarLin2012,FeiZatNag2014}. 

Extension to two-electron emission faces conceptional difficulties as to the identification of the relevant physical observables \cite{KheIvaBra2011}. 
Up to now timing information on double ionization has been indirectly extracted from spectral information by inferring from the two-electron energy and angular distribution the release time into the continuum \cite{LauBac2003,FeiNagPaz2009,PalResMcc2009,CamFisKre2012, BerKueJoh2012, PfeCirSmo2011}. 
Temporal correlations in the two-photon double ionization process could be investigated by varying the duration of the ionizing pulse (``poor-man's'' pump-probe \cite{FeiNagPaz2009}) or, e.g., by an XUV-pump XUV-probe setup studied by Palacios \etal \cite{PalResMcc2009b} where interference structures between spectrally overlapping constituents allow a reconstruction of the time elapsed between two photoabsorption events.  
For one-photon double ionization (OPDI) Emmanouilidou \etal proposed a classical two-electron streaking model  \cite{EmmStaCor2010,PriStaEmm2011,PriStaCor2012} and first timing measurements employing the RABBIT technique have been very recently reported for the OPDI of xenon \cite{ManGueArn2014}. 

\begin{figure}[tb]
  \centering
  \includegraphics[width=\linewidth]{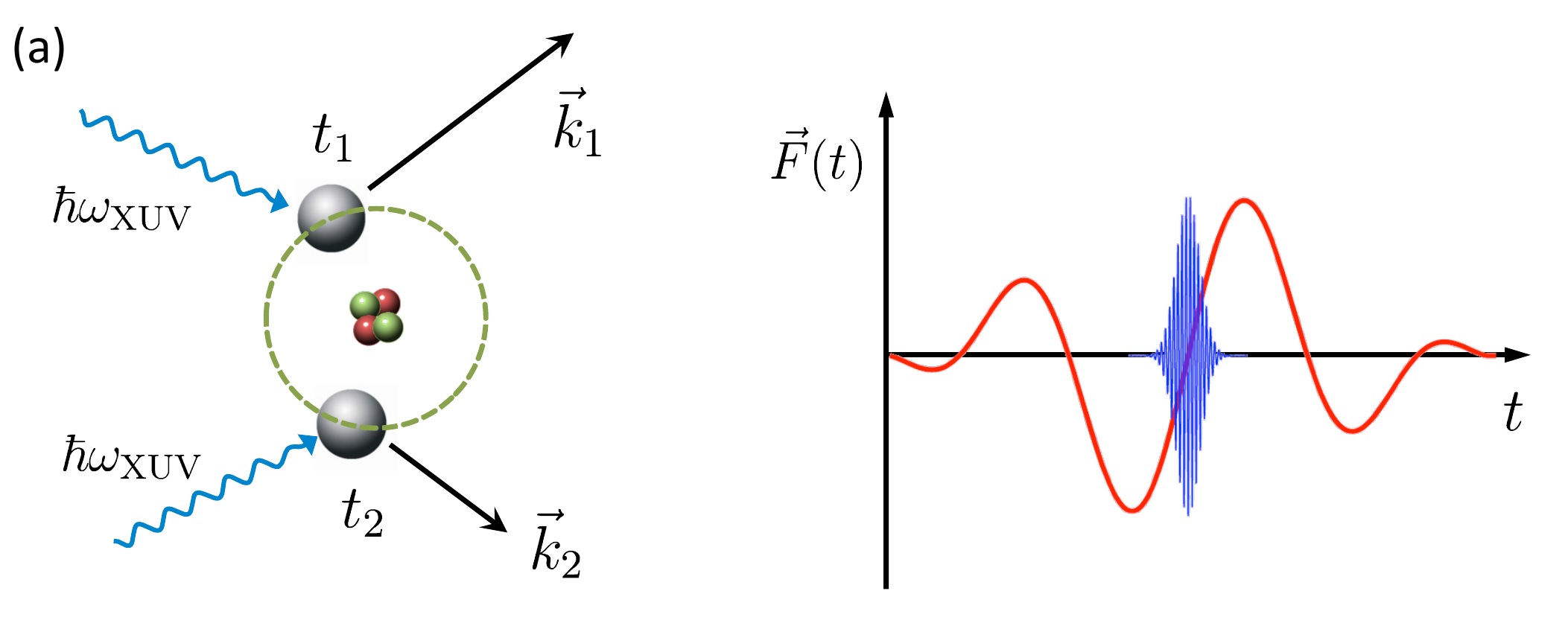}
  \includegraphics[width=\linewidth]{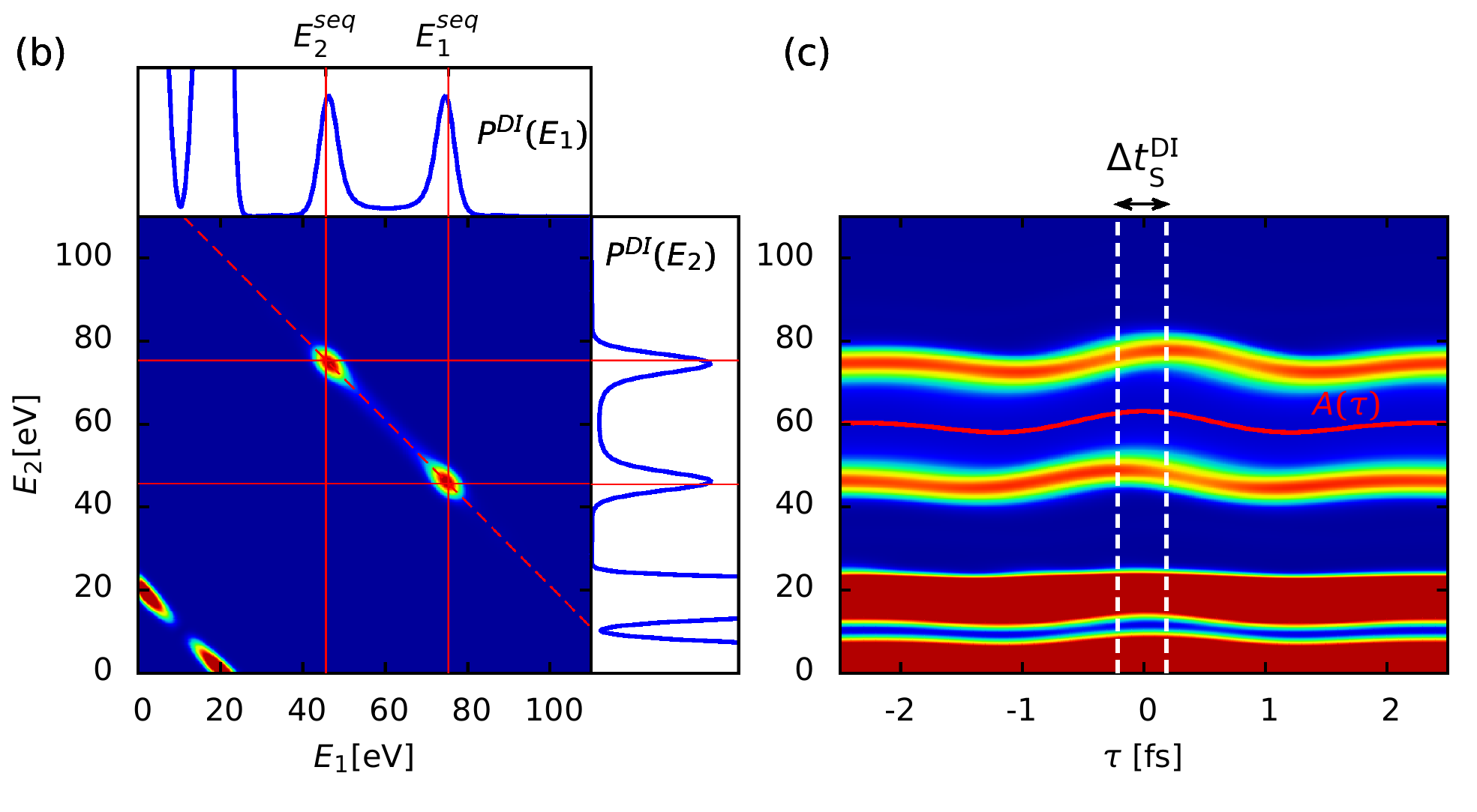}
  \caption{(a) Two-photon double ionization (TPDI) of helium by an attosecond XUV pulse (blue) in the presence of an IR streaking field (red), schematically. (b) Joint two-electron energy distribution $P^\mathrm{DI}(E_1,E_2)$ for TPDI with $\expval{\hw}=100\ev$ and a pulse duration $\tpxuv$ of $500\as$, emission back-to-back along the polarization direction ($\theta_1=0^\circ$, $\theta_2=180^\circ$). The panels above and on the right show the singly-differential energy distribution $P^\mathrm{DI}(E)$ after tracing out the energy of one electron. (c) Streaking spectrogram from the integrated spectra $P^\mathrm{DI}(E)$ in (b) at different delay times $\tau$ between the ionizing XUV pulse and the probing IR field ($\lambda\subscr{IR}=800\nm$, $\iir=4\cdot10^{11}\Wcm$, sine-squared envelope with a total duration of $6\fs$).
   } 
  \label{fig:schematics_tpdi_streaking}
\end{figure}

In this contribution we present a fully ab-initio simulation of a different two-electron process, the two-photon double ionization (TPDI) of helium (\autoref{fig:schematics_tpdi_streaking}a). 
This fundamental three-body Coulomb process has been the focus of a large number of studies in the spectral domain  (see \cite{PalResMcc2009,IshMid2005,FeiNagPaz2008,PazFeiNag2011,LauBac2003,HorMorRes2007, NikLam2007, FouHamAnt2010} and references therein), investigating the correlated energy and angular distribution of the fragments. 
Here we investigate for the first time the fully time-resolved TPDI triggered by an attosecond XUV pulse and probed by an IR streaking field. 
We show that time-resolved TPDI opens up the opportunity to explore the time ordering underlying time-dependent quantum dynamics as an accessible physical observable. 

In the energy domain (and for long XUV pulses), it has become customary to distinguish the so-called sequential (S) regime for $ \hw\subscr{XUV}>I_2=54.4\ev$ from the non-sequential (NS) regime for $(I_1+I_2)/2=39.5\ev \leq \hw\subscr{XUV} \leq 54.4\ev$, where $I_{1,2}$ are the first (second) ionization potential of helium. 
The borderline between the sequential and nonsequential  ionization is given by the binding energy $I_2$ of the most deeply bound electron of the singly ionized helium, $\Hep(1s)$. 
For photon energies above $I_2$, each electron can be ejected by one photon independent of the proximity to  and energy sharing with the other electron. 
For ultrashort pulses with $\tpxuv$ in the few-hundred attosecond regime, where the Fourier width of the pulse $\Delta \omega\subscr{XUV}\sim1/\tpxuv$ becomes comparable to the correlation energy, this distinction between sequential and non-sequential ionization becomes blurred \cite{LauBac2003,FeiPazNag2009}. 
In this regime, the TPDI is influenced by strong spatio-temporal correlation of the two-electron wavepacket irrespective of the mean frequency $\expval{\omega\subscr{XUV}}$ of the pulse. 
Real-time observation of TPDI monitored by streaking allows to inquire into the sequentiality of the emission process and the time interval between the two emissions. 

To lowest non-vanishing order perturbation theory, TPDI is given by the second-order transition matrix element 
\begin{equation}\label{eq:tdpt_transamp0}
  a_{i\to f}^{(2)} =  -\Int_{-\infty}^{\infty}\!\dt_1 \Int_{-\infty}^{t_1}\!\dt_2 
\bra{\psi_f}V_I(t_1)V_I(t_2)\ket{\psi_i}
\end{equation}
between the initial state $\ket{\psi_i}$ taken in the following to be the fully correlated He ground state and the final state $\ket{\psi_f}=\ket{\psi(\vec p_1,\vec p_2)}$ of two continuum electrons with asymptotic momenta $\vec p_1$ and $\vec p_2$ and energy $E\subscr{tot}=\sum_i{p_i^2/2}$. 
The perturbation operator in the interaction representation is given in length gauge by
\begin{equation}\label{eq:int_op}
V_I(t) = \ue^{iH_0t}\sum_{i=1}^2\vec r_i \vec F\subscr{XUV}(t) \ue^{-iH_0t}\, ,
\end{equation}
where $\vec F\subscr{XUV}(t)=F_0\exp{(-\ln 4 t^2/\tpxuv^2)}\cos(\omega\subscr{XUV}t)\hat{z}$
is the linearly polarized attosecond XUV pulse and $H_0$ is the atomic Hamiltonian. 
\autoref{eq:tdpt_transamp0} has explicitly built-in time ordering, $t_1>t_2$. 
The formation of the intermediate wavepacket $\sim V_I(t_2)\ket{\psi_i}$ by a single action of the perturbation on the initial state causing the ejection of the first electron precedes that of the wavepacket $\sim V_I(t_1)V_I(t_2)\ket{\psi_i}$ formed by the second action of the perturbation which contains a component that eventually converges towards TPDI as $t_f\rightarrow \infty$. 
The question is then posed: is such temporal sequence of events as implied by time-ordered perturbation theory physically observable even though \autoref{eq:tdpt_transamp0} represents a coherent superposition of all events without an intervening projective measurement of the intermediate state. 
We address this question with the help of a fully ab-initio solution of the time-dependent \Schro equation (TDSE) for helium in its full dimension (for details about the method see \cite{FeiNagPaz2008, SchFeiNag2011}) in the presence of the ionizing XUV field $\vec F\subscr{XUV}(t)$ and the streaking IR field $\vec F\subscr{IR}(t)$. 
The probing field is kept moderately weak with intensities $\iir\lesssim 10^{12}\Wcm$ in order to preclude unwanted ionization by the probe itself. 
While the simulation is fully non-perturbative, perturbation theory (\autoref{eq:tdpt_transamp0}) provides a useful guide for interpreting the results. 
We will demonstrate that the time-ordering underlying \autoref{eq:tdpt_transamp0} becomes visible and experimentally accessible.

The joint two-electron energy distribution for TPDI by a $500\as$ XUV field with mean photonenergy $\expval{\hw\subscr{XUV}}=100\ev$ (in the spectroscopically sequential regime) displays two distinct peaks (\autoref{fig:schematics_tpdi_streaking}b) near the energies $E_{1,2}=\expval{\hw\subscr{XUV}}-I_{1,2}$, the widths of which are governed by the Fourier width of the pulse and are also influenced by correlation effects (see \cite{IshMid2005,BarWanBur2006, FeiPazNag2009,PalHorRes2010} and references therein). 
Since the electrons are well separated in momentum (and energy) they can be easily separately traced in the same streaking spectrogram (see \autoref{fig:schematics_tpdi_streaking}c) providing a clear example for the simultaneous observation for the ``absolute'' time shift of each electron relative to the time zero, the time of the peak of the ionizing field $F\subscr{XUV}(t)$, as well as the emission time interval between the two electrons. 
This relative emission delay is so large (of the order of $\sim100\as$) that it becomes directly visible in the spectrogram without the need for a sophisticated retrieval algorithm. 
We note parenthetically that the low-energy portion ($E_{1,2}\leq20\ev$) in the joint energy distribution \autoref{fig:schematics_tpdi_streaking}b represents OPDI of helium well separated from TPDI. 
Timing information contained in the spectrogram for OPDI (\autoref{fig:schematics_tpdi_streaking}c) will be discussed elsewhere \cite{PazNagBur2014a}. 
In this contribution, we focus on the TPDI process for which already in the reduced one-electron spectra (\ie without measuring the two electrons in coincidence) the streaking delay can be easily extracted. 

\begin{figure}[bth]
  \centering
  \includegraphics[width=0.75\linewidth]{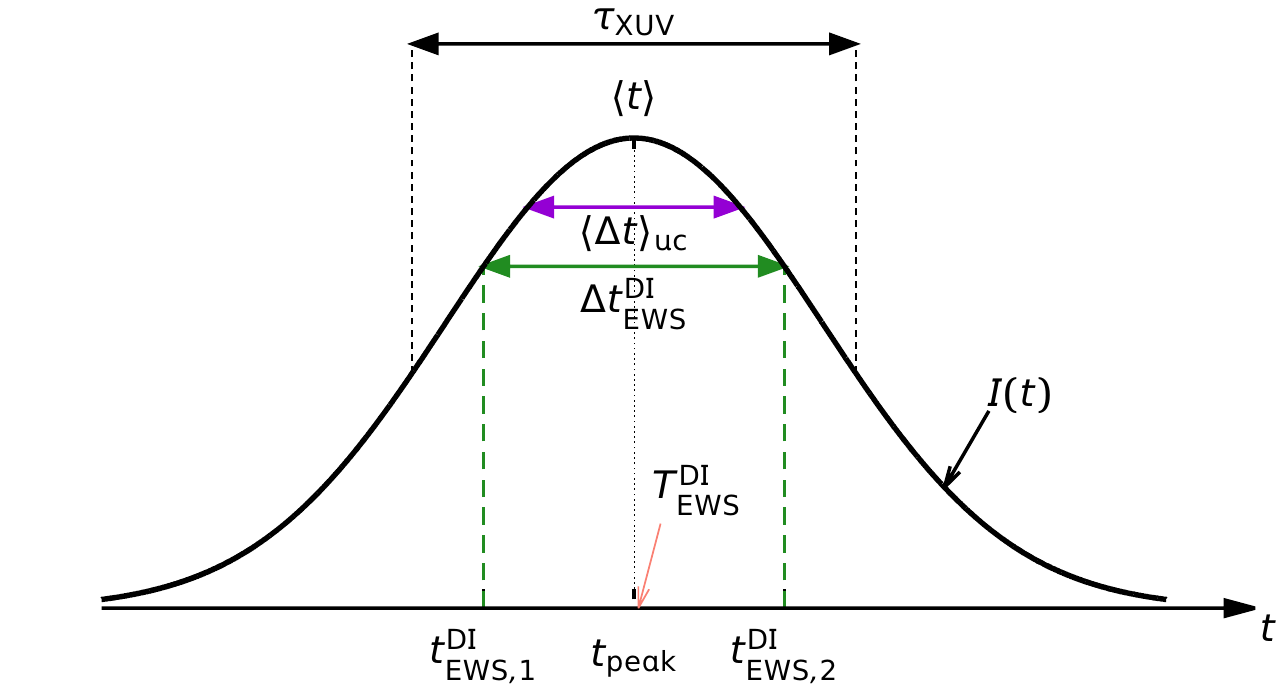}
  \caption{
  Illustration of time observables for TPDI relative to the intensity envelope $I(t)$ of the attosecond XUV pulse centered around $\expval{t}\!=\!t\subscr{peak}\!=\!0$.
   The relative emission delay between the two electrons is given by $\Delta \tewsdi=\tewsdione-\tewsditwo$. Also shown is the estimate of the relative emission delay predicted for independent uncorrelated (uc) emission events $\Delta t\subscr{uc}$. 
   } 
  \label{fig:schematics_timedelays}
\end{figure}

Identification and extraction of the relevant dynamical timing information of the two-electron wavepacket (\autoref{fig:schematics_timedelays}) is obviously more challenging than for single electron emission \cite{KheIvaBra2011} in view of the multi-dimensional nature of the final state. 
The individual one-electron EWS time shifts in the double ionization event denoted in the following by $\tewsdii \, (j=1,2)$, 
are measured relative to time zero, \ie the peak of the ionizing XUV intensity envelope $I(t)$ (\autoref{fig:schematics_timedelays}). 
Thus, a positive time shift signifies a delay or emission after the peak while a negative time shift corresponds to an advance or emission before the peak. 
While, on average, the time of absorption of a single photon may coincide with the peak of the pulse (assuming a temporally symmetric pulse shape) for a photoionization process involving two photons, deviations from this time zero are to be expected. 
Typically, one photon will be absorbed before and one after the peak (\autoref{fig:schematics_timedelays}). 
This information is encoded in the EWS times $\tewsdii$. 

The accurate determination of EWS time delays \cite{Eis1948,Wig1955, Smi1960,deCNus2002,PazNagBur2013} is not straightforward since the asymptotic scattering states are unknown. 
We therefore extract the EWS delay numerically by separately solving the TDSE for photoionization by the XUV pulse in the absence of the probing IR field, taking the energy derivative of the phase of the wavepacket (\ie its group delay) propagated to a large time $t\subscr{f}$, and subtracting the free propagation phase, $-Et\subscr{f}$ \cite{NagPazFei2012}. 
Thus, the EWS time delay for an electron with energy $E_1$ and a fixed energy $E_2$ of the other electron and fixed emission angles $\theta_1$ and $\theta_2$, emitted in TPDI follows as
\begin{multline}\label{eq:tewsdi_e1e2}
\tewsdione(E_1 | E_2,\theta_1,\theta_2)=\\
\left. \frac{\partial}{\partial E_1'}\arg\left[c^{DI}(E_1',E_2,\theta_1,\theta_2,t\subscr{f})+E_1't\subscr{f}\right]
\right|_{E_1'=E_1}
\end{multline}
where $c^{DI}(E_1,E_2,\theta_1,\theta_2,t\subscr{f})$ is the double ionization amplitude in coplanar geometry ($\phi_1\!=\!\phi_2\!=\!0$) calculated by projection of the propagated wavefunction $\psi(\vec{r}_1,\vec{r}_2,t\subscr{f})$ onto a product of uncorrelated Coulomb functions with $Z=2$ at a time $t\subscr{f}$ well after the conclusion of the XUV pulse (for the accuracy of this method see \cite{FeiNagPaz2008}). 

In addition to these ``absolute'' one-electron delays relative to the peak time of the XUV pulse, also collective two-electron time shifts can be deduced 
(\autoref{fig:schematics_timedelays}): 
the time interval between the two emission events or relative emission delay 
\begin{multline}\label{eq:tewsdi_de}
\Delta\tewsdi(\DE)=\\
\tewsdione(E_1 | E_2,\theta_1,\theta_2)-\tewsditwo(E_2 | E_1, \theta_1,\theta_2)
\end{multline}
 and the joint two-electron emission time delay
\begin{multline}\label{eq:tewsdi_etot}
\Tewsdi(\Etot)=\\
\frac{1}{2}\left[\tewsdione(E_1 | E_2,\theta_1,\theta_2)+\tewsditwo(E_2 | E_1,\theta_1,\theta_2)\right]
\end{multline}
with the energy difference $\DE\!=\!E_1\!-\!E_2$  and the total energy $\Etot\!=\!E_1\!+\!E_2$. 
$\Delta\tewsdi$ is negative when the electron with energy $E_1$ is emitted before the electron with energy $E_2$ (since in this case $\tewsdione < \tewsditwo$) and positive when the time-ordering between the two electrons is reversed (\autoref{fig:schematics_timedelays}).
The joint two-electron emission time delay $\Tewsdi$ (\autoref{eq:tewsdi_etot}), on the other hand, gives the mean delay of the collective two-electron wave packet. 

These time shifts will depend, in general, on the emission angle of the two outgoing particles. 
We will focus in the remainder on the back-to-back emission ($\theta_1=0^\circ, \theta_2=180^\circ$, \autoref{fig:schematics_tpdi_streaking}b,c) for which the interpretation of the streaking spectrogram becomes particularly simple and which also promises the highest experimental count rates as it is the most probable configuration. 

\begin{figure}[tbp]
  \centering
  \includegraphics[width=\linewidth]{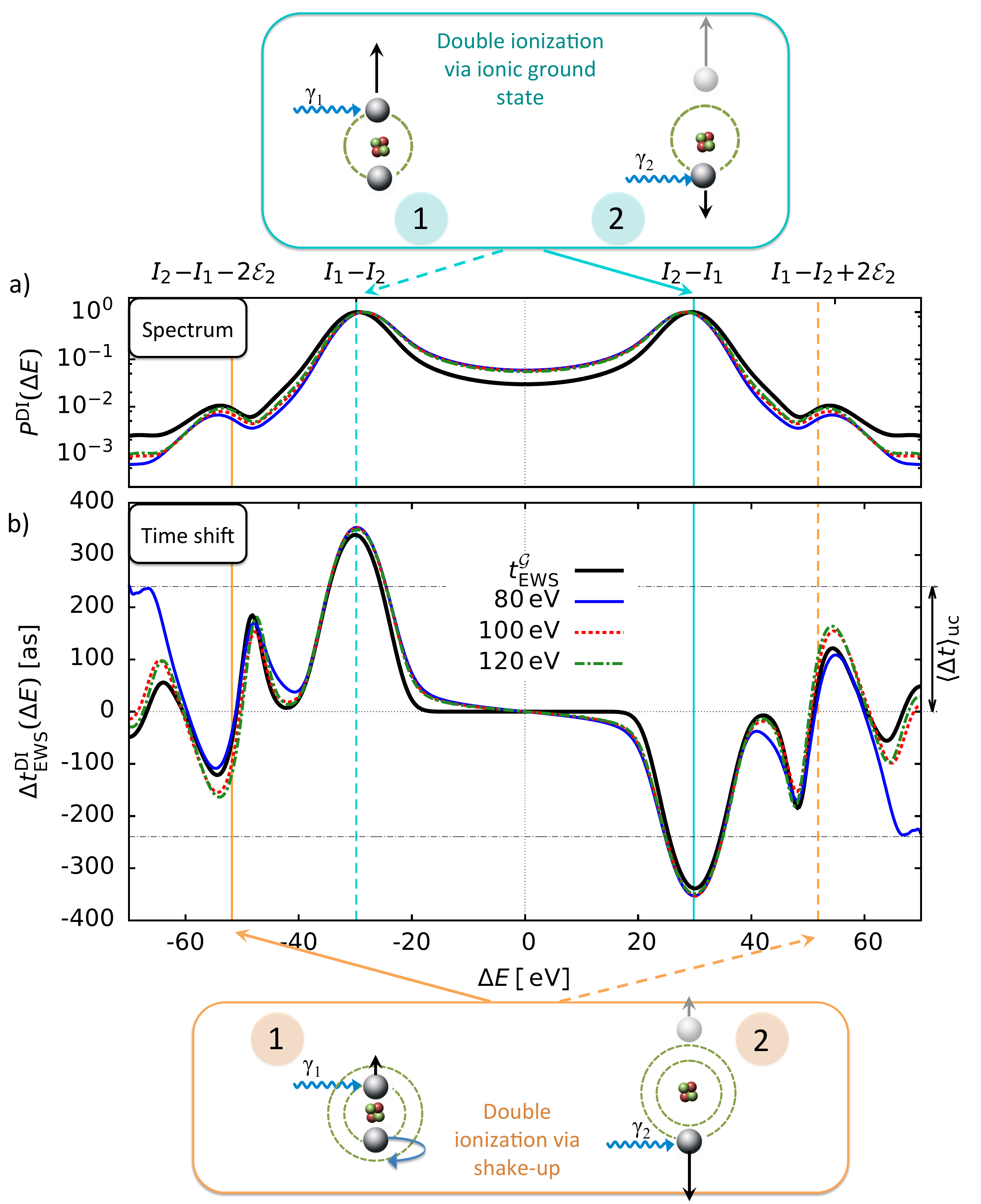}
  \caption{(a) Spectrum $P^\mathrm{DI}(\Delta E,0^\circ,180^\circ)$ and (b) emission time interval $\Delta\tewsdi(\Delta E,0^\circ,180^\circ)$, at constant total energy $\Etot=2\hw-I_1-I_2$, and back-to-back emission for different energies of the ionizing XUV pulse, $\hw\subscr{XUV}=$ 80, 100, $120\ev$.
  The Gaussian pulse has a duration $\tpxuv\!=\!500\as$ and $I=10^{13}\Wcm$. 
  The spectral positions of the peaks for sequential ionization in the limit of $\tpxuv\rightarrow\infty$ are indicated by the vertical blue (direct) and orange (shake-up) lines.
  The horizontal black dashed lines denote the time interval $\expval{\Delta t}\subscr{uc}$ (\autoref{eq:tuc}) predicted for two uncorrelated and statistically independent emission events for the given XUV pulse.
  Spectrally averaging $\Delta\tewsdi$ over the direct sequential peaks yields $\expval{\Delta t}\subscr{uc}$ to within $\sim3\as$. 
   } 
  \label{fig:tpdi_ewsde_500as}
\end{figure}

The relative emission delay $\Delta\tewsdi$ is found to be a nearly universal function of the energy difference $\DE$ while being only weakly dependent on the total energy $\Etot\!=\!2\hw\subscr{XUV}\!-\!I_1\!-\! I_2$ (\autoref{fig:tpdi_ewsde_500as}b) and thus on the XUV pulse energy.
This behaviour follows from the fact that for TPDI the spectral  (\autoref{fig:tpdi_ewsde_500as}a) and temporal  (\autoref{fig:tpdi_ewsde_500as}b)  behaviour of the two-photon wave packet is largely determined by the so-called
\emph{shape function} (\autoref{fig:tpdi_ewsde_500as}, black line) \cite{PazFeiNag2011} (see Appendix, \autoref{eq:td_pert_2nd_transamp_2_a} and \autoref{eq:td_pert_2nd_transamp_2_b}). 
The pronounced dip (\autoref{fig:tpdi_ewsde_500as}b) in the relative emission delay at $\Delta E = 30 \ev$ (vertical blue line) to $\tewsdi \sim -350\as$, corresponding to the ``sequential'' energy sharing $\Delta E =E_1-E_2\sim (\hw-I_1)-(\hw-I_2)= I_2-I_1$, unambiguously establishes that the faster, highly energetic electron is, indeed, released much earlier than the slower electron directly confirming the notion of sequential emission in the time domain: 
the ejection of the first (fast) electron with energy $E_1$ and $\theta_1\!=\!=0^\circ$ from He leaves a (near) on-shell intermediate state $\Hep(1s)$ behind from which the second (slow) electron with energy $E_2$ is emitted about $350\as$ later predominantly near $\theta_2\!=\!180^\circ$.
While the double ionization yield $P^\mathrm{DI}$ (\autoref{fig:tpdi_ewsde_500as}a) is symmetric with respect to $\Delta E$ due to the indistinguishability of the two electrons, the relative time delay $\Delta\tewsdi$ (\autoref{fig:tpdi_ewsde_500as}b) is antisymmetric, as the two cases $E_1 > E_2$ and $E_1 < E_2$  imply the opposite time ordering.
For energy differences far from on-shell intermediate states, in particular near $\Delta E\!=\!0$, the emission time interval is drastically reduced to a few attoseconds 
directly highlighting the fact that strong spatio-temporal correlations are a prerequisite in order to facilitate the required large energy sharing between the electrons for emission with small $\DE$. 
In this energy region the ionization process is, in fact, \emph{nonsequential} despite the high photon energy in the nominally ``sequential'' regime and the yields scale linearly with the pulse duration \cite{FeiPazNag2009}. 
Remarkably, the time order is preserved when the ejection of the first electron is accompanied by the formation of an intermediate shake-up state $\Hep(n\!=\!2)$ (vertical orange lines). 
Since now the roles of the fast and slow electrons are interchanged, the relative emission delay features a dip at negative values of $\DE=E_1-E_2\sim \left[\hw-(I_1+\varepsilon\subscr{n=2})\right] - \left[\hw-(I_2-\varepsilon\subscr{n=2})\right]=  I_2-I_1-2\varepsilon\subscr{n=2}\sim -50\ev$. 
Note that the position of the dip is slightly shifted and distorted by a  
dynamical Fano-``resonance-like'' lineshape resulting from the interference between the shake-up channel and the quasi-nonsequential contribution from the ground-state channel with an intermediate $\Hep(1s)$ state. 
For longer pulse durations the Fano profile for shake-up converges to a (inverted) Lorentzian profile located exactly at the energy position $\DE=\pm(I_2-I_1-2\varepsilon_n)$. 

\begin{figure}[tbh]
  \centering
  \includegraphics[width=\linewidth]{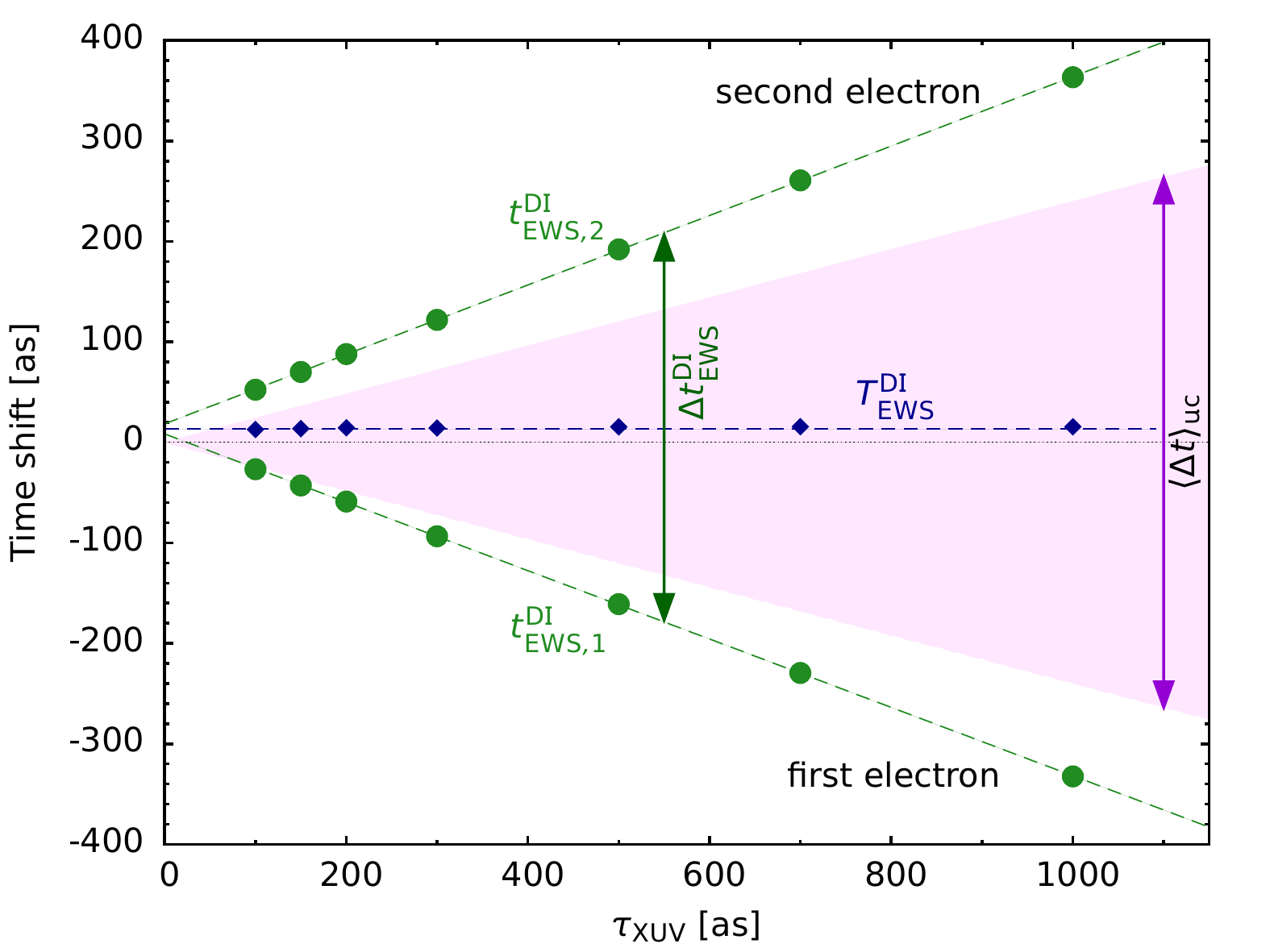}
  \caption{TPDI time shifts (green dots) $\tewsdii (j\!=\!1,2)$ as a function of the pulse duration $\tpxuv$ for $\hw=80\ev$ and back-to-back emission of the two-electrons ($\theta_1\!=\!0^\circ,\theta_2\!=\!180^\circ$) evaluated at the sequential peaks. 
The dashed line indicates the approximately linear scaling of $\tewsdii$ with the pulse duration $\tpxuv$. The purple shaded area indicates the uncorrelated mean escape delay $\dtuc$ (\autoref{eq:tuc}). 
  The joint two-electron emission time $\Tewsdi$ (\autoref{eq:tewsdi_etot}) is shown by the dark-blue diamonds.
   } 
  \label{fig:tpdi_ews_180degdurfixE}
\end{figure}

It is instructive to compare the exact time interval $\Delta \tewsdi$ with the mean time interval $\expval{\Delta t}\subscr{uc}$ predicted for two uncorrelated and statistically independent emission events with the probability density for each event proportional to the intensity of the XUV pulse, $I(t)$, 
\begin{equation}\label{eq:tuc}
\expval{\Delta t}\subscr{uc}=\tpxuv/\sqrt{\pi\ln{4}}\approx 0.479\tpxuv\, .
\end{equation}
Near the dips (or peaks) signifying sequential emission through an on-shell intermediate state, $\Delta \tewsdi$ is drastically enhanced compared to \autoref{eq:tuc} (\autoref{fig:tpdi_ewsde_500as}b). 
The linear scaling with the pulse duration (\autoref{eq:tuc}) also holds true for the absolute and relative non-perturbative quantum mechanical EWS delays $\tewsdii$ (\autoref{eq:tewsdi_e1e2}) and $\Delta\tewsdi$ (\autoref{eq:tewsdi_de}), see \autoref{fig:tpdi_ews_180degdurfixE}.
In contrast, the joint two-electron emission time delay $\Tewsdi$, signifying the mean time delay in the formation of the outgoing two-electron wavepacket relative to time zero 
is independent of the pulse duration (\autoref{fig:tpdi_ews_180degdurfixE}) but yields a constant value $\Tewsdi \approx  15\as$.
Remarkably, the extrapolation of $\tewsdii$ to the limit $\tpxuv=0$ corresponding to the limit of impulsive ionization by a broad-band pulse yields a small but finite time delay coinciding with the
joint two-electron delay $\Tewsdi$ (\autoref{eq:tewsdi_etot}) for finite pulse duration (\autoref{fig:tpdi_ews_180degdurfixE}). 

We show now that the two-electron time delays and the time-ordering of the sequential emission process become observable in attosecond streaking experiments. 
Extraction of the intrinsic time shifts for the two-electron observables of TPDI from streaking spectrograms (see \autoref{fig:schematics_tpdi_streaking}c) requires the generalization of the mapping between streaking times $\tst$ and intrinsic atomic time delays $\tews$  \cite{SchFieKar2010,PazNagBur2013,KheIva2010} to the case of two-photon double ionization. 
For \emph{one}-photon \emph{single} ionization, the streaking delay $\tS$ is extracted from the fit of the final momentum modulation to the vector potential
\begin{equation}
 \label{eq:mom_shift}
 \Delta p(t)=-A(t-\tst) \, ,
\end{equation}
as derived from the strong-field approximation (SFA) \cite{ItaQueYud2002,KitMilScr2002}.
If the SFA were exact, the streaking time shifts $\tS$ would correspond to the intrinsic atomic time delays $\tews$ \cite{SchFieKar2010}.
However, realistic TDSE simulations beyond SFA have shown that the long-range Coulomb potential gives rise to an additional Coulomb-laser coupling term $\tCLC$ \cite{NagPazFei2011,ZhaThu2010,PazNagBur2013}.
Accordingly, 
\begin{equation}
 \label{eq:streaking_decomposition}
 \tS = \tEWS + \tCLC \, .
\end{equation}
Additional dipole-laser coupling contributions present for strongly polarizable systems \cite{NagPazFei2011,PazFeiNag2012} can be safely neglected in the present case.

\begin{figure}[tbh]
  \centering
  \includegraphics[width=\linewidth]{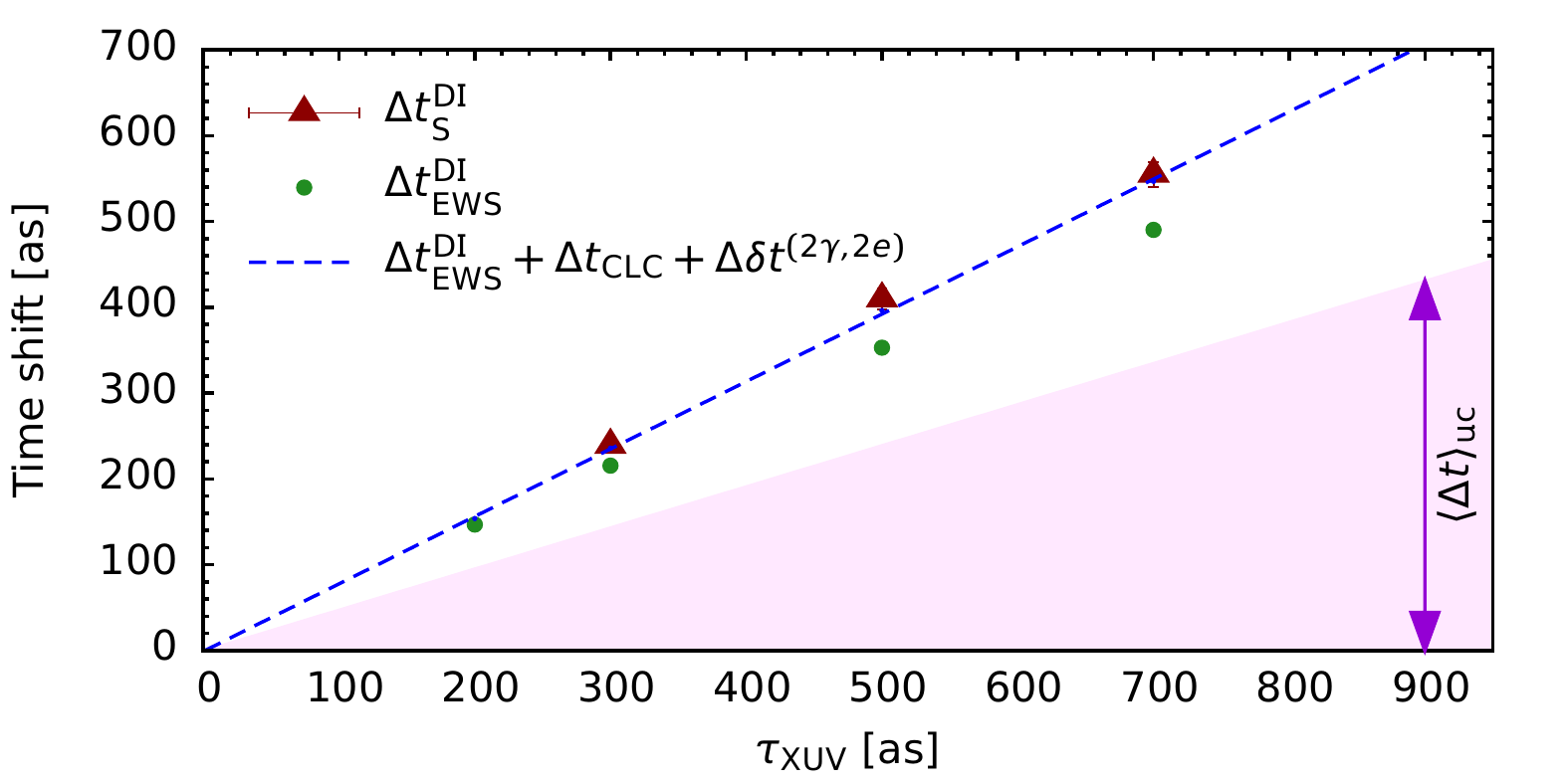}
  \caption{Time interval between the two subsequent photoemission events in TPDI as a function of the pulse duration $\tpxuv$ for $\hw=80\ev$ and back-to-back emission of the two-electrons ($\theta_1\!=\!0^\circ,\theta_2\!=\!180^\circ$). 
  The relative streaking time shifts $\Delta\tst^\mathrm{DI}$ (red triangles) extracted from a streaking spectrogram as in \autoref{fig:schematics_tpdi_streaking} for $\iir=10^{10}\Wcm$ and $\lambda\subscr{IR}=800\nm$ are compared with the right hand side of \autoref{eq:ts_di_corr}, 
  the sum of $\Delta \tewsdi$, $\Delta \tclc$, and $\Delta \tSSFA$ (blue dashed line). 
  Separately shown is the contribution $\Delta \tewsdi$ (green dots).  
   } 
  \label{fig:tpdi_ts}
\end{figure}

For double ionization, the IR streaking field leads to a modification of the final momenta in the $p_1$ -- $p_2$ plane and likewise of the final energies in the $E_1$ -- $E_2$ plane (\autoref{fig:schematics_tpdi_streaking}b). The analysis of the streaked two-electron spectra is most conveniently performed after integration over one energy leading to the spectrogram \autoref{fig:schematics_tpdi_streaking}c. 
However, this nontrivial mapping results in an additional time shift specific for TPDI, $\tSSFA$. 
Accordingly, the streaking time shift of the $j^\mathrm{th}$ electron, $\tSdii \, (j=1,2)$, observed in TPDI of the fully Coulomb-interacting system reads
\begin{equation}\label{eq:ts_di_corr}
\tSdii=\tewsdii+\tclci+\tSSFAi \, .
\end{equation}
\autoref{eq:ts_di_corr} represents the generalization of the relationship between streaking time shifts and EWS delays for TPDI. 
The additional correction term in \autoref{eq:ts_di_corr}, $\tSSFAi$, specific to TPDI, can be determined by comparison with a (numerical)  two-electron SFA calculation (see Appendix).
Unlike for one-photon ionization, the EWS delays for (sequential) two-photon ionization 
 do not only depend on the atomic properties of the system under scrutiny (\ie the dipole matrix elements) but also on the temporal structure of the ionizing pulse. 
In our simulations we can extract the \emph{absolute} streaking time shifts $\tSdii$ (\autoref{eq:ts_di_corr}) by comparison of the streaking traces with the  IR vector potential. By contrast, the \emph{relative} streaking time shift $\Delta\tSdi = \tSdione - \tSditwo$ can be measured from the temporal offset between the two bands in the spectrograms (\autoref{fig:schematics_tpdi_streaking}c).
We have verified the relation \autoref{eq:ts_di_corr} for a wide range of XUV pulse durations (\autoref{fig:tpdi_ts}) and XUV energies. 
All terms on the right hand side of \autoref{eq:ts_di_corr} can be independently and accurately determined. 
We find excellent agreement with the ab initio simulation for $\tSdi$ (left hand side of \autoref{eq:ts_di_corr}) on the $\lesssim 10\as$ level (\autoref{fig:tpdi_ts}). 
The residual error is of the order of the uncertainty in the extraction of $\tS$ for the two-electron wavepacket. 
\autoref{fig:tpdi_ts} also clearly demonstrates that the time delay between the two emission events, $\Delta \tewsdi$, and, thus, the time ordering of emission can be accurately extracted from attosecond streaking traces. 

The experimental challenge for the realization of the proposed protocol lies in the separation of the comparably weak double ionization signal from the dominant single ionization channel. 
The higher energetic peak at $\sim75\ev$ in \autoref{fig:schematics_tpdi_streaking}c overlaps with the single ionization signal. 
The latter is, however a factor $1.5\times10^3$ larger than the TPDI signal for an XUV intensity of $10^{13}\Wcm$ (15 times for $I\subscr{XUV}=10^{15}\Wcm$). 
Therefore, coincident detection of the doubly charged ion $\Hepp$ is the prerequisite to discriminate against the single ionization channel. 
However, coincidence detection of the two electrons is \emph{not} required for the present protocol. 
We are therefore confident that with the advances in the generation of more intense XUV pulses an experimental realization of the proposed scheme will become possible in the near future. 

In summary, the present ab initio streaking simulations for two-photon double ionization show that atomic time delays, in particular the time interval elapsed between the two photoemission events can be observed in real time with an accuracy better than $10\as$. 
The notion of (non)sequential photoemission originally developed in the realm of spectroscopy can now be directly verified in the time domain for ultrashort pulses. 
Moreover, the concept of time ordering underlying time-dependent perturbation theory is accessible in measurements of sequential photoemission without compromising the coherence of the underlying time evolution.

We thank Johannes Feist for his work on the helium code and for fruitful discussions in the early stage of this work. This work was supported by the FWF-Austria, Grant No.\ P21141-N16, P23359-N16, SFB 041 (ViCoM) and SFB 049 (Next-Lite), the COST Action CM1204 (XLIC), and in part by the National Science Foundation through XSEDE resources provided by NICS and TACC under Grant TG-PHY090031. The computational results presented have also been achieved in part using the Vienna Scientific Cluster (VSC). RP acknowledges support by the TU Vienna Doctoral Program Functional Matter.\\

\appendix
\section*{Appendix A: Streaking of two-photon time delays}
In this appendix we provide technical details underlying the determination of 
streaking and EWS time delays for 
 wavepackets created in photoionization by \emph{two} photons presented in the main text. 
 With the help of lowest-order time-dependent perturbation theory (TDPT)
we show that the wavepacket group delay contains two contributions: one stemming from the dipole transition matrix element which carries information about the atomic structure and another one from the time structure of the ionizing XUV pulse. 

We start from the second-order TDPT amplitude (\autoref{eq:tdpt_transamp0}) which can be factorized as
\begin{equation} \label{eq:td_pert_2nd_transamp_2_a}
  a_{i\to f}^{(2)} = -\sumint{n} \braOket{f}{\op\mu}{n} \braOket{n}{\op\mu}{i} \G\left[E_{f},E_{n},E_i,F\subscr{XUV}(t)\right]
\end{equation}
with
\begin{multline} \label{eq:td_pert_2nd_transamp_2_b}
   \G \left[\Efn,\Eni,F_\mathrm{XUV}(t)\right] = \\
   \Int_{t_0}^{\infty}\!\dt_1 \Int_{t_0}^{t_1}\!\dt_2 e^{i E_{fn} t_1} e^{i E_{ni} t_2} F\subscr{XUV}(t_1) F\subscr{XUV}(t_2) \, .
\end{multline}

The so-called \emph{shape function} $\G$ \cite{PazFeiNag2011,PalResMcc2009} (which would in the one-photon case reduce to the simple Fourier transform of the field) is a functional of $F\subscr{XUV}(t)$ and a function of the energy differences $E_{fn}=E_f-E_n$ and $E_{ni}=E_n-E_i$, with $E_i=E_0$, $E_n=E[\Hep(n)]+E_1$, and $E_f=E_1+E_2$ ($E_1=k_1^2/2$, $E_2=E_2^2/2$) for TPDI of helium. 
The sum over intermediate states $\sum_n\ket{n}\bra{n}$ contains virtual and (near) on-shell singly ionized states.

Considering for notational simplicity just one single intermediate state $n$ in \autoref{eq:td_pert_2nd_transamp_2_a}, the DI EWS delay for the electron $j$ emitted with energy $E_j$ [
$\tewsdii$ (\autoref{eq:tewsdi_e1e2})] can be approximated within TDPT as a sum
\begin{align}\label{eq:arg_tif}
 \frac{\partial}{\partial E_j}\arg{a_{i\rightarrow f}^{(2)}} = & 
\frac{\partial}{\partial E_j} \arg{\braOket{\psi_n}{\op \mu}{\psi_i}} + 
  \frac{\partial}{\partial E_j}\arg{\braOket{\psi_f}{\op \mu}{\psi_n}} \nonumber \\ 
 &+  \frac{\partial}{\partial E_j}\arg{\G\left[\Efn,\Eni,F_\mathrm{XUV}(t)\right]}\\
=&\sum_{m=1}^{2} \tewssijk(E_j) + \tGtj(E_1,E_2) \label{eq:arg_tif2}
 \, .
\end{align}
In \autoref{eq:arg_tif}, $\braOket{\psi_n}{\op \mu}{\psi_i}$ and $\braOket{\psi_f}{\op \mu}{\psi_n}$ are the (one-photon) dipole matrixelements connecting the initial with the intermediate state and the intermediate with the final state, respectively. 
Their spectral phase derivatives correspond to the one-photon EWS delays $\tewssijk$ for the two ionization steps $(m=1,2)$ resulting from the absorption of the two photons $\gamma_m$.
The spectral derivative $\tGt(E_1,E_2)$ of the shape function $\G$ gives rise to an additional contribution, to the time delay specific to the two-photon ionization process. 
Thus, the DI EWS delay can be decomposed into (i) contributions $\tewssijk$ that stem from the one-photon matrix elements of the two subsequent ionization events of He and $\Hep$, and (ii) a term $\tGt$ that is given by the shape function of second-order TDPT which only depends of the temporal structure of the XUV pulse and the ionization potentials of the system.
This term carries the information on the delay between the absorption time of the two photons.

In the limit of a purely sequential ionization passing through an on-shell intermediate state of $\Hep$, $\braOket{\psi_n}{\op \mu}{\psi_i}$ reduces to the matrix element of single ionization of $\He$, \ie emitting the first electron with energy $E_1$ whereas the second electron remains bound, and $\braOket{\psi_f}{\op \mu}{\psi_n}$ represents the emission of the second electron with $E_2$ from the $\Hep$ ion. 
Furthermore, assuming the two ionization processes to be uncorrelated, the transition amplitudes can be approximated by $\braOket{n\ell,\vec{k}_1}{\op \mu}{1s^2}$ and $\braOket{\vec{k}_2,\vec{k}_1}{\op \mu}{n\ell,\vec{k}_1}$ so that to the spectral derivative in \autoref{eq:arg_tif} with respect to $E_1 (E_2)$ only the first (second) matrix element contributes. 
By comparing with the numerically exact expression $\tewsdi$ [\autoref{eq:tewsdi_e1e2}] we find that for high photon energies above the double-ionization threshold ($\hw\subscr{XUV} \gtrsim 80\ev$), the one-photon timeshifts $\tewssijk$ in \autoref{eq:arg_tif2} can be approximated by the corresponding 
Coulomb EWS delay for $Z=2$ given by the Coulomb phase $\sigma_\ell$, $\tEWSC(E,Z\!=\!2,l\!=\!1)=\ddE \sigma_\ell(E,Z\!=\!2)$ \cite{PazNagBur2013} evaluated at energies $E_1$ and $E_2$ with errors smaller than 3 attoseconds. 
Likewise, the collective two-electron emission time delay $\Tewsdi$ (\autoref{eq:tewsdi_etot}, \autoref{fig:tpdi_ews_180degdurfixE}) can be decomposed into the EWS delays for the individual, independent ionization steps $\tewssijk$ (He $\to \Hep$ and $\Hep \to \Hepp$) of about $10\as$ and a remaining contribution of about $5\as$ due to electron-electron correlations in the ionization process.

Interrogation of the TPDI process by the IR streaking field maps the delay time $\tewsdii$ (\autoref{eq:arg_tif}) onto the streaking time shift $\tSdii$. 
Both the one-photon contributions $\tewssijk$ and the two-photon contribution $\tGt$ acquire additional probe-field induced time shifts that are additive. 
While the one-photon contributions are modified by the Coulomb-laser coupling time $\tCLC$, the two-photon term is corrected by the $\tSSFA$ streaking field term. 
Accordingly, 
\begin{multline}\label{eq:tsdig}
\tSdii\simeq\sum_{m=1}^{2} \tewssijk(E_j) + \tclci(E_j) + \\
\tGtj(E_1,E_2) + \tSSFAi(E_j)
\end{multline}
which, without the $n=1$ restriction, results in \autoref{eq:ts_di_corr} of the main text, 
\begin{equation}\label{eq:tsdig}
\tSdii= \tewsdii(E_j) + \tclci(E_j) + \tSSFAi(E_j) \, .
\end{equation}
The TPDI streaking correction $\tSSFA$ can be determined invoking the strong-field approximation that also underlies the original identification of $\tst$ (\autoref{eq:mom_shift}) \cite{YakBamScr2005}.

Accordingly, we calculate a two-electron SFA reference streaking spectrogram using \autoref{eq:td_pert_2nd_transamp_2_a} for which we switch off the atomic contribution to the time delay by setting all transition matrix elements equal to unity.
The presence of the streaking field is non-perturbatively included through Volkov energy phases,
\begin{multline}\label{eq:2o_sfa1}
a_{i\rightarrow f}^{\mathrm{\scriptstyle{DI,S}}}(\vec{p}_1,\vec{p}_2) =\\
-\Int_{-\infty}^{t_f}\dt_1
e^{{\left[i\left(\frac{p_2^2}{2}t_1+\vec{p_2}\vec{\alpha}\subscr{IR}(t_1)+\frac{\mathcal{A}\subscr{IR}(t_1)}{2} +I_2 t_1\right)\right]}}
\vec{F}\subscr{XUV}(t_1) \times \\
\Int_{-\infty}^{t_1}\dt_2
e^{{\left[i\left(\frac{p_1^2}{2}t_2+\vec{p_1}\vec{\alpha}\subscr{IR}(t_2)+\frac{\mathcal{A}\subscr{IR}(t_2)}{2} +I_1 t_2\right)\right]}}
\vec{F}\subscr{XUV}(t_2)
\end{multline}
with
\begin{equation}\label{eq:A_defs}
\vec{\alpha}(t)=\Int_{-\infty}^{t}\vec{A}(t')\dt'\, ,\quad  
\mathcal{A}(t)=\Int_{-\infty}^{t}\vec{A}^2(t')\dt' \, . 
\end{equation}
The resulting one-electron streaking spectrum after integration over the energy of the second electron is, analogously to \autoref{eq:mom_shift},
\begin{equation}
 \label{eq:mom_shift_TPDI}
 \Delta p(t)=-A(t-\tS) = -A\left(t-\tGt-\tSSFA\right) \, .
\end{equation}
The TPDI-specific additional streaking time shift $\tSSFA$ can thus be determined by subtracting from the calculated SFA streaking time shift $\tS$ the independently determined EWS time delay associated with the shape function for TPDI, $\tGt$,
\begin{equation}
 \label{eq:TPDI_correction}
\tSSFAi(E_j) = \tS(E_j) - \tGt(E_j) \, .
\end{equation}
The correction term $\tSSFAi$ depends on the intensity of the probing IR field as well as on the electron energy for IR intensities $>\!10^{10}\Wcm$. 

\newpage

\bibliographystyle{jpbjo}
\bibliography{citeulike_atto_cleaned}
\end{document}